\documentclass[pre,preprint,english,aps,prb,a4paper,floatfix]{revtex4}

\usepackage{graphics}
\usepackage{epsfig}

\usepackage{amssymb}
\usepackage{bm}
\usepackage{color}

\begin{document}




\title{Two-dimensional nematics in bulk and confined geometries}

\author{D. de las Heras}
\address{Theoretische Physik II, Physikalisches Institut, Universit\"at Bayreuth, D-95440 Bayreuth, Germany}
\author{Y. Mart\'{\i}nez-Rat\'on}
\address{Grupo Interdisciplinar de Sistemas Complejos (GISC), Departamento de Matem\'aticas, Escuela Polit\'ecnica Superior, Universidad Carlos III de Madrid, Avenida de la Universidad 30, 28911 Legan\'es, Madrid, Spain}
\author{L. Mederos}
\address{Instituto de Ciencia de Materiales de Madrid, Consejo Superior de Investigaciones Cient\'{\i}ficas, Campus de Cantoblanco, 28049 Madrid, Spain}
\author{E. Velasco}
\address{Departamento de F\'{\i}sica Te\'orica de la Materia Condensada, Universidad Aut\'onoma de Madrid, 28049 Madrid, Spain}

\begin{abstract}
Two-dimensional nematics possess peculiar properties that have been studied recently using computer
simulation and various theoretical models. Here we review our own contribution to the field using
density-functional theory, and 
present some preliminary simulation results on confined two-dimensional nematics.  
First we discuss the possible stable bulk phases and phase diagrams and the relation between 
phases and particle geometry. We then explore the adsorption properties on a single substrate and the
confinement effects that arise when the fluid is confined  
between parallel walls. Next, confinement in circular cavities is presented; this geometry
allows us to measure some properties of the simplest defects that arise in
two-dimensional nematics. Finally, preliminary Monte Carlo simulation results of confined nematics in 
circular geometry are shown.
\end{abstract}

\maketitle

%



\section{Introduction}
\label{Introduction}

The nematic phase arises in liquids composed of sufficiently anisotropic particles (molecules, colloidal
particles, macroscopic grains) from the isotropic phase when the density or concentration of particles
(or temperature in thermal liquids) is above (below) some particular value \cite{deGennes}. In nematics, particles are oriented on 
average along a common direction $\bm{n}$, called the {\it director}, but retain their positional disorder. The gain in 
orientational order that occurs when the isotropic phase transforms into the nematic phase proceeds via a
phase transition which, in three dimensions, can be shown to be of first order, although weak. In this respect, two-dimensional 
(2D) nematics are peculiar, since
symmetry restrictions on the orientational order parameter, discussed most conveniently 
in the context of a Landau theory, allow for a more
general nature of the phase transition: it can be of either first-order or continuous. 

As is the case with three-dimensional nematics, in 2D point defects play an important role. Point defects are
points in the plane where the nematic director field $\hat{\bm n}(x,y)$ is not defined; there exists an associated region around the
point defect where the director field
is highly distorted from the uniform configuration $\hat{\bm n}=$ const. The density field $\rho(x,y)$ is also affected.
Another important aspect of nematics is that they are very sensitive to the presence of surfaces, so that they 
can be used to orient the director in some specific directions. Finally, the broken rotational symmetry of
nematics brings about an elasticity which, in two dimensions, is restricted to two independent elastic modes
controlled by their corresponding elastic constants $k_1$ and $k_3$. These three elements, defects, surfaces and elasticity,
often compete together to give the final configuration of the nematic, which in many cases is highly nontrivial.

Similar to many other 2D systems, the nematic state in 2D presents anomalously large thermal
fluctuations which result in a highly fluctuating nematic director. In case the free energy can be described by a 
one-elastic-constant, Frank-type elastic model,
\begin{eqnarray}
F_e=\frac{1}{2}\int_Ad^2r\left\{k_1\left(\nabla\cdot\hat{\bm n}\right)^2+k_3\left|\nabla\times\hat{\bm n}\right|^2\right\}=
\frac{1}{2}\int_Ad^2r k\left|\nabla\theta\right|^2,
\label{elastic}
\end{eqnarray}
[where $\hat{\bm n}=\left(\cos{\theta},\sin{\theta}\right)$], the fluctuations in $\theta$, the director tilt angle, 
depend on the number of particles $N$ as $\left<\theta^2\right>\sim 
\log{N}$, with a vanishing order parameter in the thermodynamic limit, $q_1=\left<\cos{2\theta}\right>\sim N^{-kT/2\pi k}$, 
and an orientational correlation function $g_n(r)=\left<\cos{n\theta(r)}\right>\sim r^{-n^2kT/2\pi k}$ which decays 
algebraically rather than presenting long-range order. All of these properties imply that, strictly speaking, 
the ordered nematic phase does not exist in the thermodynamic limit, although the dependence is slow and even large
nematic samples, or confined nematics, will be well ordered.

\section{Bulk behaviour of hard models}
\label{Hard models}

As shown by Onsager \cite{Onsager}, overlap or exclusion interactions alone (hard particles) can explain the formation of the ordered nematic
phase in two and three dimensions. In these systems, one defines particles with a specific shape, a convex body
in the present work, and 
associates a potential energy to a pair of particles such that the energy is zero if particles do not overlap
and infinity otherwise. Overlapping configurations are thus discarded from the partition
function, and the average energy is zero. Order is controlled by the entropy: in the case of nematics, 
the orientational entropy, which favours disorder, is more than compensated by the exclusion configurational entropy,
which increases as particles align along the director. This happens above some particular particle number
density.

\begin{figure}
\epsfig{file=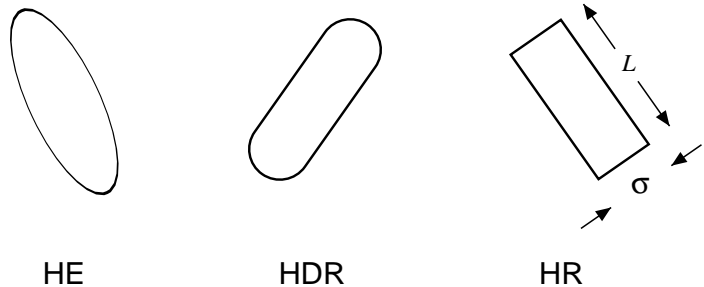,width=4.0in}
\caption{Popular models for convex hard particles in two dimensions. HE: hard ellipses. HDR: hard discorectangles.
HR: hard rectangles. In the case of HR the length $L$ and width $\sigma$ of the particle are indicated. For HDR, the
total length of the particle is $L+\sigma$.}
\label{fig1}
\end{figure}

Popular hard models for nematic formation in two dimensions are shown in Fig. \ref{fig1}.
They include: hard ellipses (HE),
hard discorectangles (HDR), and hard rectangles (HR). The thermodynamics and phase behaviour of these 
models have been analysed using computer simulations.
In the case of the HE and HDR models two liquid phases are observed: the isotropic phase (I), which exists at low
density, and the uniaxial nematic phase (N$_u$), at higher densities. In the N$_u$ phase the continuous rotational symmetry
of the I phase is broken and a definite average orientation appears in the fluid. The I-N$_u$ transition is believed to be continuous
and of the Kosterlitz-Thouless type \cite{Bates}. 

\begin{figure}
\begin{center}
\includegraphics[width=4.0in]{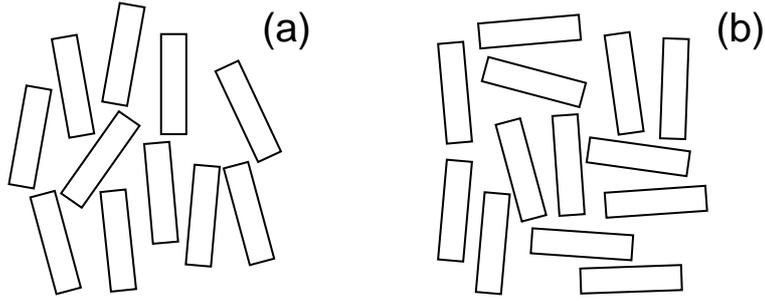}
\caption{\label{fig2} Typical particle configurations in (a) the uniaxial nematic phase N$_u$, and (b) the tetratic 
N$_t$ phase.}
\end{center}
\end{figure}

The HR model contains sharp corners and is close to being non-convex. This shape 
favours a new, exotic type of nematic, called the tetratic phase (N$_t$) \cite{Schlacken}. 
In the tetratic phase four equivalent directions,
specified by two equivalent directors, arise. Fig. \ref{fig2}
is a schematic of a typical particle configuration in the two nematic 
phases. Also, the I-N transition in the HR fluid is highly non-trivial since its nature depends in a complicated way on the
particle aspect ratio $\kappa=L/\sigma$ ($L$ being the rectangle length and $\sigma$ the width).
Hints of the phase behaviour found for these fluids might also be observed in granular quasi-two-dimensional fluids made of 
anisotropic particles and subject to vertical motion \cite{hindues}.

\subsection{One-component HR and HDR fluids}

A deep understanding of the I-N transition in fluids of hard particles is obtained by focusing on the excluded volume
$v_{\hbox{\tiny exc}}$: the presence of one particle creates a surrounding region from which a second particle is excluded. 
The available volume for the latter, $V^{\prime}=V-v_{\hbox{\tiny exc}}$, where $V$ is the system volume, depends on the relative 
orientation of the two particles, and is maximised when parallel. In this two-particle view of the problem
the excluded volume (area in two dimensions) becomes a central quantity. 

The excluded area as a function of relative angle $\phi$ of the HR model differs in a crucial way from that of the HDR model.
These functions can be obtained analytically:
\begin{eqnarray}
&&v_{\hbox{\tiny exc}}^{\hbox{\tiny HDR}}(\phi)=4L\sigma+\pi\sigma^2+L^2\left|\sin{\phi}\right|,\nonumber\\
\nonumber\\
&&v_{\hbox{\tiny exc}}^{\hbox{\tiny HR}}(\phi)=\left(L^2+\sigma^2\right)\left|\sin{\phi}\right|+2L\sigma
\left(1+\left|\cos{\phi}\right|\right).
\end{eqnarray}
These functions exhibit a minimum at $\phi=0$ and $\pi$ but, in addition,
$v_{\hbox{\tiny exc}}^{\hbox{\tiny HR}}(\phi)$ presents a secondary minimum at $\phi=\pi/2$ which, when the aspect ratio $\kappa$
is sufficiently small, the fluid uses to stabilise the tetratic phase. 

\begin{figure}
\begin{center}
\includegraphics[width=3.0in]{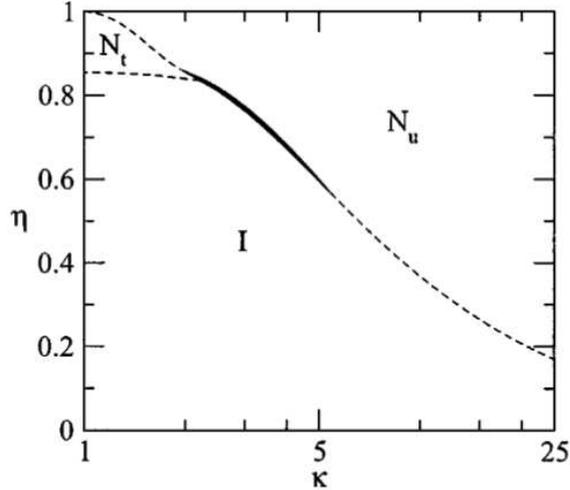}
\caption{\label{fig3} Phase diagram of HR in the $\eta$--$\kappa$ plane as obtained from SPT, showing regions
of stability of isotropic (I), uniaxial nematic (N$_u$) and tetratic (N$_t$) phases. Dashed curves: continuous phase
transitions. Shaded region: first-order phase transition. Points where the continuous transition changes to first order
are tricritical points.}
\end{center}
\end{figure}

A fruitful theory to explore the thermodynamics and structure of the isotropic and nematic phases is the Scaled-Particle Theory (SPT)
\cite{Schlacken,Cotter,Scott}.
It is a density-functional theory: a free-energy density $\Phi[h]$ is written in terms of the orientational distribution function
$h(\phi)$. The latter gives the probability density that a particle is oriented with 
angle $\phi$. Note that $\int_0^{\pi}d\phi h(\phi)=1$.
The free energy density $\Phi[h]=-Ts[h]$, where $T$ is temperature and $s$ entropy density, is split in ideal
\begin{eqnarray}
\beta\Phi_{\hbox{\tiny id}}[h]=\rho\left(\log{\eta}-1+\int_0^{\pi}d\phi h(\phi)\log{\left[\pi h(\phi)\right]}\right)
\end{eqnarray}
and excess
\begin{eqnarray}
\beta\Phi_{\hbox{\tiny ex}}[h]=\rho\left[-\log{\left(1-\eta\right)}+\frac{\eta}{1-\eta}S_0\right]
\end{eqnarray}
parts, where $\eta$ is the packing fraction, defined as
$\eta=\rho L\sigma=NL\sigma/A$ with $A$ the system area, and $\beta=1/kT$. Here
\begin{eqnarray}
S_0=\frac{1}{2v}\int_0^{\pi}d\phi\int_0^{\pi}d\phi^{\prime}h(\phi)v_{\hbox{\tiny exc}}(\phi-\phi^{\prime})h(\phi^{\prime})-1,
\end{eqnarray}
where $v$ is the particle area.
By numerically minimising $\Phi[h]$ with respect to $h(\phi)$ (which is most easily done in Fourier space)
\cite{Geometry}, one obtains the
thermodynamic functions and the equilibrium orientational structure. Order parameters can then be defined,
\begin{eqnarray}
q_1=\int_0^{\pi}d\phi h(\phi) \cos{2\phi},\hspace{0.4cm}q_2=\int_0^{\pi}d\phi h(\phi) \cos{4\phi}
\end{eqnarray}
which measure respectively the usual nematic order and the tetratic order. The three fluid phases involved have the
following order parameters: $q_1=q_2=0$ (I); $q_1\ne 0$ and $q_2\ne 0$ (N$_u$); and $q_1=0$, $q_2\ne 0$ (N$_t$).

The SPT phase diagram of the HR model in the $\eta$ vs. 
aspect ratio $\kappa$ plane is shown in Fig. \ref{fig3}. At high aspect ratios
there is a continuous I-N$_u$ transition that becomes of first order as particles become less elongated.
The tetratic phase is stable for $\kappa<2.21$, and is stabilised from the I phase at $\eta\simeq 0.85$ via a continuous
phase transition. Note that this value of $\eta$ is expected to be well above the transition to phases with spatial order
(either smectic, columnar or crystal). This point is discussed in Sec. \ref{FMT}.

Finally, the fluid phase diagrams of the HE and HDR models, not shown, are much simpler, since only the I and N$_u$ phases are stable, with a continuous
transition in the whole range of $\kappa$ (note that the phase diagrams of all models, HE, HDR and HR,
tend to be similar for large $\kappa$, and in fact 
identical in the limit $\kappa\to\infty$). This result agrees with simulations \cite{Eppenga,Cuesta,Bates}. The only 
non-uniform phase present at high density is the crystal phase \cite{Cuesta,Bates}.

\subsection{Beyond two-particle correlations}

SPT only takes account of two-particle correlations. But higher-order correlations must be important in the stabilisation
of the N$_t$ phase. This is because, in typical N$_t$-like arrangements and in order to obtain more optimised packings, 
particles tend to order in small clusters that involve a few particles pointing in perpendicular directions. 
Because of this effect,
higher-order correlations should extend the island of N$_t$ stability to aspect ratios higher than $2.62$ and packing
fractions lower than $0.85$. 

In \cite{3body} we made an attempt to include three-particle correlations in a way that 
(i) incorporates three-body effects, and (ii) reduces to the SPT model when three-body correlations are switched off. 
The latter are 
included by means of the third virial coefficient, which is computed numerically as a function of the two order parameters
$q_1$ and $q_2$. In effect, the new theory predicts a larger island of tetratic stability, with the aspect ratio 
where the tetratic phase ceases to be stable shifting to $3.23$, and the line defining the I-N$_t$ transition moving down
to the range $\eta=0.70$--$0.75$. Further inclusion of higher-order correlations will no doubt refine these results,
although the formulation of new theories is difficult for lack of approximations for the higher-order virial
coefficients and the intrinsic difficulties to deal with these coefficients. 

An alternative approach was followed in \cite{Clustering}, where emphasis was focused on the strong clustering observed in the
HR fluid. Monte Carlo simulations show that, because of particle shape and low dimensionality, particles tend to 
align parallel to their neighbours thus inducing the formation of large and persistent clusters that dominate the structure
of the fluid. The clusters consist of parallel hard rectangles, side by side, say $n$ in number, which form a ``super-rectangle''
of length $n\sigma$ and width $L$. With this fact in mind, the fluid can be viewed as a polydispersed mixture of
super-rectangles of different lengths $n\sigma$ \cite{Clustering}, and the extension of SPT to mixtures can be applied. The only
unknown quantity in the problem is the distribution of cluster sizes, given by $f(n)$. 
Assuming this function to be exponential, $e^{-\lambda n}$
(which follows from a chemical mass-action law, 
an assumption which is supported by simulation), we obtain $\lambda$ from MC 
simulation and predict the phase behaviour. The tetratic phase boundary can be shown to shift quite substantially as a 
result of clustering effects, thus increasing the stability range of the N$_t$ phase.

\subsection{Spatially-ordered phases}
\label{FMT}

\begin{figure}
\begin{center}
\includegraphics[width=3.0in]{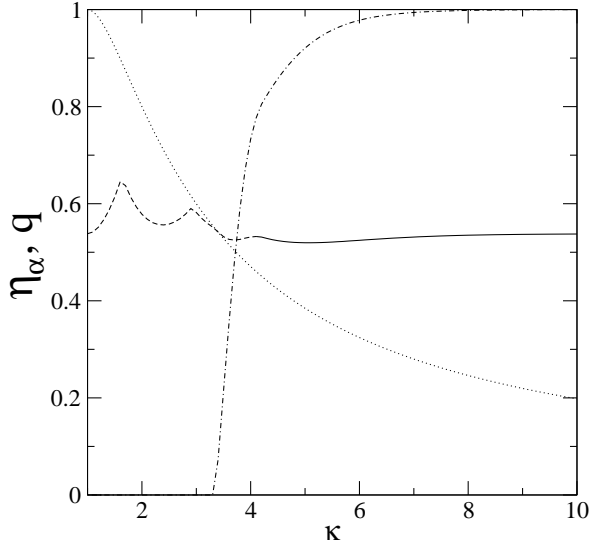}
\caption{\label{FMT_fig} Phase diagram of the FMT model in the $\eta$--$\kappa$ plane using the Zwanzig approximation.
Dotted curve: isotropic-nematic transition. Dashed and continuous curves: bifurcation lines to spatially-ordered phases.
Dashed-dotted line: nematic order parameter $q_1$ at the bifurcation line to the spatially-ordered phases.
See text for details.}
\end{center}
\end{figure}

A more convenient framework than SPT to study phases with spatial order in hard-particle models is fundamental-measure theory 
(FMT). This theory, originally proposed by Rosenfeld \cite{Rosenfeld} for hard spheres, has been extended to other particle shapes; in particular, there
is a version for rectangles in 2D. By formulating a theory for two species, each associated to one of two mutually 
perpendicular directions, one can treat fluids with (restricted) orientational order. This is the Zwanzig approximation, where one
deals with two densities, $\rho_x({\bm r})$ and $\rho_y({\bm r})$, with ${\bm r}=(x,y)$ and expresses the local excess free energy 
density $\Phi_{\hbox{\tiny exc}}({\bf r})$ as
\begin{eqnarray}
\beta\Phi_{\hbox{\tiny exc}}({\bf r}) =-n_0({\bf r})\log{\left[1-n_2({\bf r})\right]}+\frac{n_{1x}({\bf r})n_{1y}({\bf r})}{1-n_2({\bf r})},
\end{eqnarray}
in terms of averaged densities
\begin{eqnarray}
n_{\alpha}({\bm r})=\sum_{\nu=x,y}\left[\rho_{\nu}\star\omega_{\nu}^{(\alpha)}\right]({\bm r}),\hspace{0.4cm}
\alpha=\{0,1x,1y,2\},
\end{eqnarray}
which are obtained from convolutions of the densities with one-particle geometrical measures
$\omega_{\nu}^{(\alpha)}({\bf r})$ (see \cite{Yuri} for details). To $\Phi_{\hbox{\tiny exc}}({\bf r})$ we add the 
local mixing entropy density $\Phi_{\hbox{\tiny id}}({\bf r})$, 
\begin{eqnarray}
\beta\Phi_{\hbox{\tiny id}}({\bf r})=\sum_{\nu=x,y}\rho_{\nu}({\bf r})\left[\log{\rho_{\nu}}({\bf r})-1\right],
\end{eqnarray}
(which is directly related to an orientational entropy) and minimise the total free energy per unit volume 
${\cal F}/V=V^{-1}\int_V d{\bf r}\left[\Phi_{\hbox{\tiny id}}({\bf r})+\Phi_{\hbox{\tiny exc}}({\bf r})\right]$
with respect to both local densities. The local density $\rho({\bf r})$ and the uniaxial order parameter $q_1({\bf r})$ can be defined as 
$\rho({\bf r})=\rho_x({\bf r})+\rho_y({\bf r})$ and $q_1({\bf r})=[\rho_x({\bf r})-\rho_y({\bf r})]/\rho({\bf r})$, respectively. 
In Fig. \ref{FMT_fig} the phase diagram of the FMT model
in the $\eta$--$\kappa$ plane is presented. This diagram extends the calculations presented in \cite{Yuri}, where only
the case $\kappa=3$ was studied, although here we only calculate the spinodal instabilities (not coexistence calculations) to non-uniform phases. 
In this model, the I-N transition (dotted line) is always continuous (note that transition 
densities are
considerably lower in this model than in SPT, due to the restricted-orientation approximation). The continuous
line corresponds to the spinodal (bifurcation) of the nematic-smectic transition. The dashed line represents either
the nematic-columnar (when $q_1$, also represented in dashed--dotted line, is non-zero), isotropic-columnar or isotropic-plastic
solid (these two cases cannot be distinguished from each other and correspond to the case $q_1=0$).  
Note that the transition from uniform to non-uniform phases is at about $\eta\sim 0.5$ in the whole range of aspect ratios
$\kappa$; therefore, assuming that a freely-rotating model showed instabilites in the same range as in the present model, it is unlikely that a tetratic phase can truly be stabilised if it only exists, even in metastable state,
beyond this value.



\subsection{Mixtures of two-dimensional particles}

We have just seen that 2D particles possessing anisometric shape can form nematic phases. Depending on the particle
geometry the nematic phase can occur in two varieties, the uniaxial and the tetratic nematic phases, but the latter
exists only for low aspect ratios. The isotropic-nematic transitions are generally continuous, except for the
HR model in a small range of aspect ratios. An interesting question is how this scenario may change when particles with
different geometries and elongations are mixed.

\begin{figure}
\begin{center}
\includegraphics[width=2.8in]{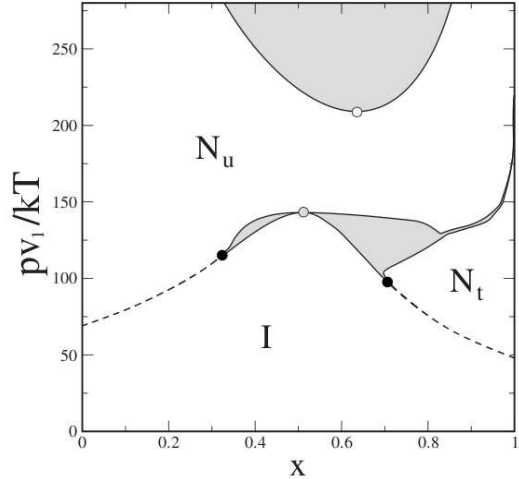}
\caption{\label{mix}Phase diagram for a HR/HDR mixture in the scaled-
pressure $pv_1/kT$ vs composition $x$ plane, where $x$ is the number fraction of the rectangles.
Values of the parameters are: for the rectangles, $\kappa_1 = L_1/\sigma_1=1.5$ and $\sigma_1 = 1$; for the discorectangles,
$\kappa_2 = (L_2+\sigma_2)/\sigma_2=2$ and same particle area as a rectangle of aspect ratio equal to 2 and unit breadth. Open circle indicates 
the critical point, and shaded circle denotes an azeotropic point. Gray areas are two-phase regions, while dashed lines are
continuous phase transitions.}
\end{center}
\end{figure}

This problem can be tackled using the same theoretical scheme as before, i.e. SPT, but extended to include a mixture
of two components \cite{Demixing,mixtures,Yuri3}, specifically HR and another component, such as hard discs, HE, HDR or 
HR with different aspect ratio. 
Here orientational ordering and demixing effects compete to give a complex phase
diagram. As expected, when rectangles are mixed, in
increasing concentration, with other particles not possessing stable tetratic order by themselves (e.g. discs,
HDR or HR of high aspect ratio), the tetratic
phase is destabilized, via a continuous or discontinuous phase transition, to uniaxial nematic or isotropic phases.
Strong demixing behaviour is observed when particle geometries of the two components are very different, or simply
aspect ratios are very different. 
As an example, Fig. \ref{mix} shows the phase diagram of a mixture of HR with aspect
ratio $\kappa_1=L_1/\sigma_1=1.5$ and $\sigma_1 = 1$, and HDR of aspect ratio 
$\kappa_2=(L_2+\sigma_2)/\sigma_2=2$ in the reduced pressure $pv_1/kT$ versus composition $x$
plane ($\sigma_2$ is chosen as explained in the caption of Fig. \ref{mix}). 
The nature of the isotropic-nematic transition 
gets quite complex depending on the mixture composition, with intervals where the transition is continuous or of first-order.  
At higher pressure a region of strong demixing appears in the uniaxial nematic region, bounded by a lower critical point.  

\subsection{Elastic constants}
\begin{figure}[h]
\begin{center}
\includegraphics[width=4.0in]{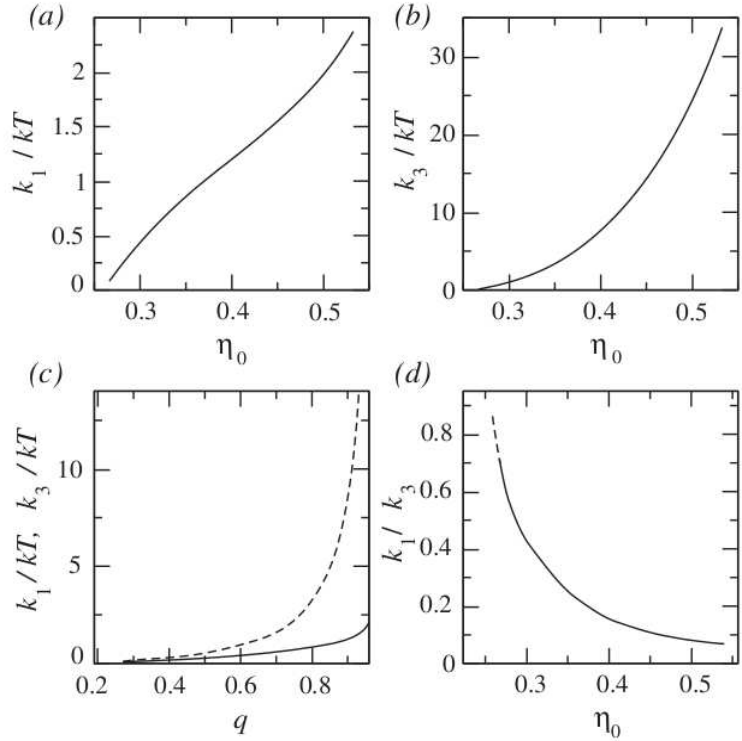}
\caption{\label{fig9} Elastic constants of a nematic fluid of HDR
particles with aspect ratio $\kappa=16$, as obtained from DFT. (a) Splay elastic constant $k_1$ as a function
of packing fraction $\eta$; (b) Bend elastic constant $k_3$ as a function of $\eta$; (c) The two elastic
constants as a function of uniaxial order parameter $q_1$; (d) Ratio of $k_1$ to $k_3$ as a function of $\eta$.}
\end{center}
\end{figure}

A two-dimensional nematic exhibits two independent elastic modes: splay and bend, controlled respectively by the
elastic constants $k_1$ and $k_3$. For a general deformation of the director field $\hat{\bm n}(x,y)$, the macroscopic elastic 
free energy can be written as in Eqn. (\ref{elastic}), and the constants appear as curvatures of the free energy.
General expressions for the elastic constants exist in terms of the correlation function of the model fluid
\cite{Stecki,Cavity1}. In the case
of SPT, the correlation function is proportional to the Mayer function (basically the excluded area), and these
integrals can be calculated numerically without much effort \cite{Cavity1}. Fig. \ref{fig9} shows the values of the elastic constants
for the HDR model with $\kappa=16$. Obviously the elastic constants are zero at the bulk transition. Note that $k_3$ is always larger 
than $k_1$ (so that bend deformations are more costly energetically than splay deformations), with their difference increasing with density. 
At the highest nematic densities their ratio is about 10; this is an important point since most analyses based on elastic theory 
assume the one-constant approximation $k_1=k_3$, an approximation which may introduce significant errors.

\subsection{Adsorption and confinement in a slit pore}

\begin{figure}
\begin{center}
\includegraphics[width=4.0in]{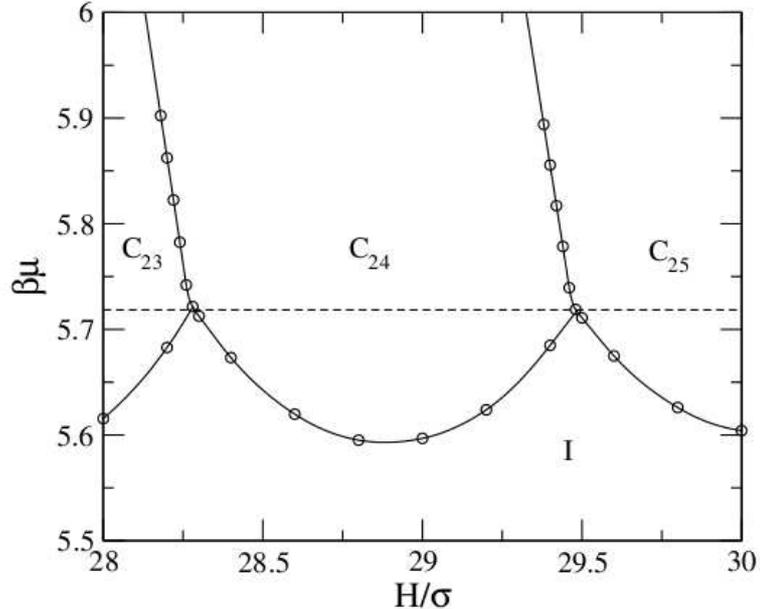}
\caption{\label{Yuri_fig} Phase diagram of HR under confinement in a slit pore in the chemical potential $\mu$
vs. pore width $H$ for aspect ratio $\kappa=3$. Lines indicate first-order transitions (circles are actual calculations).
I: isotropic phase. C: columnar phase. The subscripts for the
C labels correspond to the number of columns in the confined phase. The horizontal line is the value
of chemical potential $\mu$ for the bulk I-C transition.}
\end{center}
\end{figure}

Since the HR model exhibits a rich bulk phase diagram, one anticipates fascinating surface and adsorption properties.
In \cite{Yuri} a semiinfinite system consisting of an isotropic fluid in contact with a single (hard line) wall
was studied. The favoured particle configuration next to the wall was observed to be parallel; this orientation
propagates into the bulk material. As density is increased, complete wetting of the wall-isotropic interface by the columnar phase was found.
The theoretical model used was a density-functional theory based on the fundamental-measure functional in the 
Zwanzig approximation where, as mentioned before, particles are restricted to lie only along one of the two Cartesian 
axes. 

In the same work, the case of confinement by two such walls was studied \cite{Yuri}. 
In line with the phenomenology observed in 
studies of the same problem but in three dimensions \cite{smec,smec1}, a complex behaviour was found involving capillary columnar ordering 
and layering transitions (Fig. \ref{Yuri_fig}). The latter occur as a result of competition between the natural periodicity of 
the columnar phase and the separation between the two walls. A strong coupling between this phenomenon and that of capillary
ordering was also observed, and the complete surface phase diagram was obtained. Fig. \ref{Yuri_fig} shows a region of
such phase diagram, centred at the layering transitions involving 23 to 25 columns. The isotropic-columnar transition in the pore
is not monotonic, and oscillates depending on the commensuration between the columnar period and the pore width.


\subsection{Confinement in a circular cavity}

\begin{figure}
\begin{center}
\includegraphics[width=2.0in]{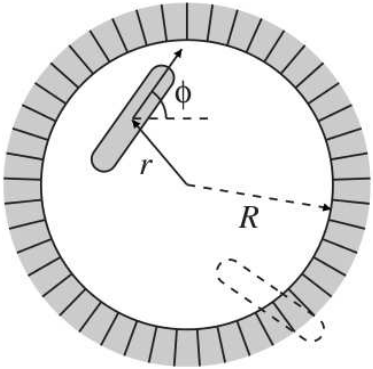}
\caption{\label{fig6} Hard discorectangle inside a hard circular wall of radius $R$.
The wall acts on the centre of mass of the particle, as indicated by the particle with dashed boundary which is
right at contact with the wall.}
\end{center}
\end{figure}

The circular geometry creates an additional geometric constraint that gives rise to new phenomenology, specifically,
to the formation of defects. There are two aspects of this problem. One is the thermodynamical aspect, which is 
important in assessing the different stable phases that arise in the cavity. The other is the structural aspect.
The two are of course intimately connected. 

The wall imposes severe restrictions as to the favoured
particle alignment next to it but, by construction, the circular geometry creates a deformation of the director
field and a corresponding elastic free-energy cost. In turn, because of topological constraints,
the deformation into a closed region generates a
defect. Again, competition arises between surface, elastic and defect free energies which gives a rich phase
behaviour as density (or chemical potential) and cavity radius are varied. On top of this, there is a
modified (capillary) isotropic-nematic transition that is strongly affected by the confinement.

In \cite{Cavity2} we used a modified Onsager theory, based on excluded-volume interactions, to analyse the thermodynamics
of the HDR fluid confined by a hard circular wall, Fig. \ref{fig6}, using particles with aspect ratio 
$\kappa=(L+\sigma)/\sigma=15$ (note that, for such a large value of aspect ratio, the results for the corresponding
HR fluid will be very similar, so our particular choice of particle geometry is not crucial). As mentioned before, it turns out that the combination HDR fluid/hard wall 
brings about parallel alignment of the nematic director. We forced the alignment to be perpendicular by making the wall
act on the particle centres of mass; the so-called `homeotropic anchoring' results, but the associated surface
free energy is not too large compared with $kT$ (weak anchoring conditions). 

\begin{figure}
\begin{center}
\includegraphics[width=3.0in]{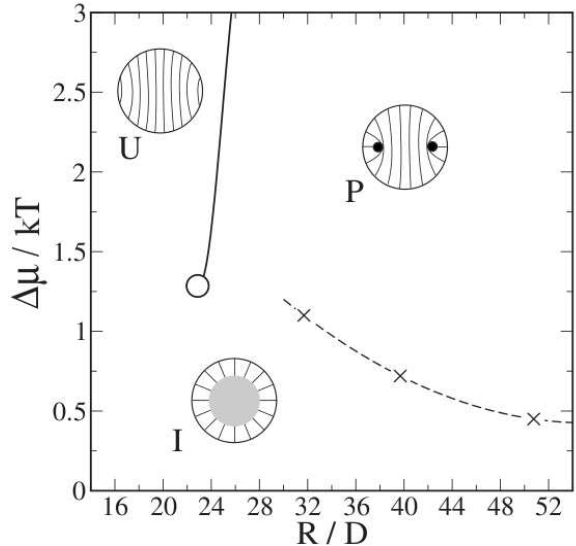}
\caption{\label{fig7} Phase diagram of a fluid of hard discorectangles with aspect ratio
$\kappa=(L+\sigma)/\sigma=15$ inside a hard circular wall in the $\Delta\mu$ vs. $R$
plane. $\Delta\mu$ is the chemical potential with respect to that of the bulk fluid, and $R$ the cavity radius. U: uniform phase.
P: polar phase. I: isotropic phase. Icons
on the right represent the director configuration in each phase. Lines are phase transitions; solid curve: first-order U-P transition,
ending in a terminal point (open circle); dashed curve: continuous I-P transition (crosses correspond to actual
calculations).}
\end{center}
\end{figure}

Under these circumstances, it is found that, for small cavity radii
$R$, the cavity is free of defects at the expense of surface free energy not being optimized, and a more or less
uniform director field results, the so-called `uniform', U phase. However, for larger cavity radius, a defect with
total topological charge $k = +1$ stabilises inside the cavity. This defect adopts one of two possible configurations: 
(i) either
a single, central `hole' of charge $k = +1$ with a size that depends on $R$ and the chemical potential 
$\Delta\mu=\mu-\mu_{\infty}$ 
($\mu_{\infty}$ being the bulk chemical potential), the so-called `isotropic' phase, I; (ii) or one consisting of two
defects of charge $k=+1/2$, of more or less constant size, located along a diameter symmetrically with respect to the
cavity centre, but with a separation that depends on $R$ and $\Delta\mu$, the `polar' phase, P. See Fig. \ref{fig7},
where the phase diagram for $\kappa=15$ in the $\Delta\mu$-$R$ plane, with the different phase boundaries, is shown.
The U and P phases are separated by a line (continuous curve in the figure) where the structure of the
fluid changes discontinuously. The line ends in a terminal point below which the two structures cannot be 
distinguished.

Computer simulations on HDR fluids confined into circular cavities \cite{Dzubiella} and on hard ellipsoids in
cilindrical containers \cite{Andrienko} indicate that a single, $k=+1$ point defect never stabilizes and that
the configuration with two $k=+1/2$ defects is always more stable. A careful
interpretation of our results supports this finding, since the I phase is in fact related to the bulk isotropic phase 
($R\to\infty$) and is stable below a line (dashed curve in the figure) which we can refer to as {\it capillary}
isotropic-nematic transition. We can then say that a nematic
confined in a circular cavity always exhibits two $k=+1/2$-point defects instead of a single $k=1$-point defect,
presumably because of the largest free energy of formation of the latter.

The location of the capillary I-N transition on the phase diagram deserves some comments. In bulk
($R\to\infty$) the transition is continuous in our mean-field model (as commented above, simulations point to a 
fluctuation-driven phase transition which cannot be correctly described by a mean-field theory). As soon as the fluid
is confined into a circular cavity, long-wavelength fluctuations are supressed and, even worse, there can be no
phase transition in the strict sense due to the finite size of the system (this is in contrast with other types
of confinement where at least one direction is infinite). The mean-field model, however, continues to exhibit
features that can be interpreted as remnants of the bulk transition. For example, the integrated order parameter
exhibits a kink which we associate with a `ghost' capillary I-N transition; this is the dashed curve plotted in
Fig. \ref{fig7}. In our calculations the curve could not be continued to smaller cavities due to numerical inaccuracies,
and therefore we are uncertain as to whether the curve continues and touches the structural transition (continuous curve) 
or else terminates in some kind of end-point.

\begin{figure}
\begin{center}
\includegraphics[width=5.0in]{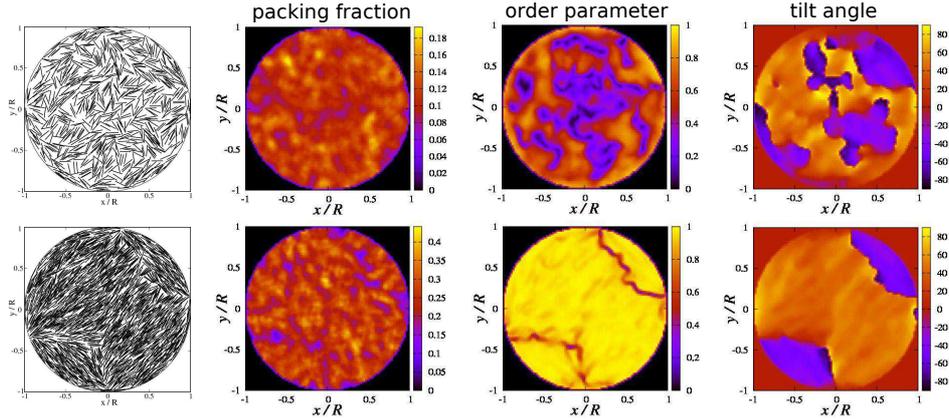}
\caption{\label{MC} Panels from left to right: particle configurations and packing fraction, uniaxial order parameter and
director tilt angle local fields. Top panels correspond to a number of particles $N=800$, while the bottom panels pertain
to the case $N=1970$. In all cases the cavity radius is $R=7.5 L$, and the particle aspect ratio is $\kappa=40$.}
\end{center}
\end{figure}

\subsection{Monte Carlo simulation of confinement in a circular cavity}

Recently we have been looking at the problem of confined nematics in 2D from the point of view of
Monte Carlo (MC) simulation. To this effect, we have used a HR model inside a circular 
cavity but, this time, the hard wall acts upon the whole particle and not just on its centre of mass,
so that the favoured particle alignment at the wall is parallel. Instead of focusing on the thermodynamics and 
phase changes in the cavity, we are interested in the structure and type of defects that occur inside the
cavity. This work is still in a preliminary stage, and we only report some representative behaviour in the r\'egime
of cavity radius $R/L<15$. Larger cavity sizes will be considered in future studies.

We consider HR with aspect ratio $\kappa=40$; we chose this value to match the value used by Galanis et al. \cite{Galanis}
in their experiments of vibrated granular rods.
We set the radius of the cavity to $R=7.5 L$, and start from a very dilute cavity with only a few particles.
When this state is equilibrated, particles are added one by one at random positions and with random
orientations, and the system is equilibrated between consecutive additions. All simulations use the
standard NVT Monte Carlo algorithm for constant number of particles and area.

Fig. \ref{MC} corresponds to the case
$N=800$ (top) and $N=1970$ (bottom). In each case the following is shown, from left to right: particle configurations
in a representative state, and equilibrium (ensemble averaged) local fields for, respectively, packing fraction, 
uniaxial nematic order parameter and director tilt angle with respect to the (fixed) $x$ axis 
(note that averages do not contain a sufficiently large number of
configurations so that the fields are a bit coarse). The case $N=800$ corresponds to an isotropic phase (note that the tilt 
angle is not significant wherever the order parameter is too low), while the 
other contains a nematic phase, with an overall high value of order parameter.

In the nematic case a remarkable structure forms inside the cavity. It consists of one large region with two other, 
smaller regions separated from the first
by wavy defect lines on which the order parameter is suppressed. The distortion of the director in each of the three
regions is kept to a minimum (i.e. particles are close to being parallel) while, at the same time, particles mostly
satisfy the orientation favoured at the wall. The director field rotates by 90$^{\circ}$ across the defect lines. 
This is the preferred particle configuration of the sample, and not one containing a central $k=+1$ defect or two 
$k=+1/2$ defects at opposite sides of the circular wall. We have checked that this structure obtained by MC is 
robust with respect to the initial configuration used to start the dynamics; for instance, one can start from a 
structure with two $k=+1/2$ defects, and the MC dynamics invariably leads to the same final structure with two defect 
lines.  

In their study of elastic and anchoring constants in quasimonolayers of vibrated steel rods,
Galanis et al. \cite{Galanis} presented results for the case $R=4.7L$ which resemble very much the defected
structure observed by us in our MC thermal simulations. We suspect that this type of structure may be
favoured for low values of the ratio $R/L$, and that the typical `polar' nematic phase with two defects inside
the cavity will be obtained for larger cavities. A study of this problem is now being investigated in our group.  








\section{Conclusions}

Despite their apparent simplicity, 2D nematics continue to present new challenges in their bulk and 
surface properties. In bulk, a new type of 2D nematic, the tetratic phase, was recently discovered using a simple SPT 
approach. This exotic phase, akin to the cubatic phase found in three-dimensional nematic fluids, could be
stable for hard rectangular particles of low aspect ratio; this is still to be seen, but strong
tetratic correlations are actually observed in normal, uniaxial phases, as shown in computer simulations
\cite{Donev}, experiments on colloidal rectangles \cite{Chaikin}, and in vibrated experiments on 
quasimonolayers of granular rods \cite{hindues}, even for aspect ratios of at least $\kappa\sim 6$.
The absence of strict long-range orientational order in 2D nematics, their strong fluctuations and
non-standard transition to the isotropic phase add difficulties in the effort to reach a deeper 
understanding of this problem; clearly, more sophisticated theories and additional computer simulations 
are needed.

On the other hand, 2D nematic are a useful testbed to understand the formation of defects in confined geometries
and also the subtle commensuration effects, combining surface, bulk, elastic and defect energies, that operate
in confined systems.\\
\\
\noindent{\bf Acknowledgements}\\
\\
\noindent
We acknowledge financial support from Comunidad Aut\'onoma de Madrid under the R\& D Program of
Activities MODELICO-CM/S2009ESP-1691, and from MINECO (Spain) under grants MOSAICO, FIS2010-22047-C01, FIS2010-22047-C04 and
FIS2010-22047-C05.

\end{document}